\documentclass[twocolumn]{revtex4}
\usepackage{graphicx}
\usepackage{times}
\begin{document}
\newcommand{\rme}{{\mathrm{e}}} 
\newcommand{\rmd}{{\mathrm{d}}}
%
\newcommand{\fig}[2]{\includegraphics[width=#1]{./figures/#2}}
\newcommand{\pfig}[2]{\parbox{#1}{\includegraphics[width=#1]{./figures/#2}}}
\newcommand{\Fig}[1]{\includegraphics[width=\columnwidth]{./figures/#1}}
\newlength{\bilderlength}
\newcommand{\bilderscale}{0.25}
\newcommand{\storebilderscale}{\bilderscale}
\newcommand{\bilderskip}{\hspace*{0.8ex}}
\newcommand{\textdiagram}[1]{%
\renewcommand{\bilderscale}{0.23}%
\diagram{#1}\renewcommand{\bilderscale}{\storebilderscale}}
\newcommand{\diagram}[1]{%
\settowidth{\bilderlength}{\bilderskip%
\includegraphics[scale=\bilderscale]{./figures/#1}\bilderskip}%
\parbox{\bilderlength}{\bilderskip%
\includegraphics[scale=\bilderscale]{./figures/#1}\bilderskip}}
\newcommand{\Diagram}[1]{%
\settowidth{\bilderlength}{%
\includegraphics[scale=\bilderscale]{./figures/#1}}%
\parbox{\bilderlength}{%
\includegraphics[scale=\bilderscale]{./figures/#1}}}

\author{Pierre Le Doussal{$^1$}, Kay J\"org Wiese{$^1$}, Elie Raphael{$^2$} and Ramin Golestanian{$^3$}}
\affiliation{{$^1$}CNRS-Laboratoire de Physique Th{\'e}orique de l'Ecole
Normale Sup{\'e}rieure, 24 rue Lhomond 75005 Paris, France \\
{$^2$}Laboratoire de Physique de la Mati\`ere
Condens\'ee, Coll\`ege de France, 11 place Marcelin Berthelot 75005 Paris, France \\
{$^3$} Institute for Advanced Studies in Basic Sciences, Zanjan
45195-159, Iran}

\title{Can Non-Linear Elasticity Explain 
Contact-Line Roughness at Depinning?}
\begin{abstract}\smallskip We examine whether cubic non-linearities,
allowed by symmetry in the elastic energy of a contact line, may
result in a different universality class at depinning.  Standard
linear elasticity predicts a roughness exponent $\zeta=1/3$ (one
loop), $\zeta=0.388\pm 0.002$ (numerics) while experiments give $\zeta
\approx 0.5$. Within functional RG we find that a non-local KPZ-type
term is generated at depinning and grows under coarse graining. A
fixed point with $\zeta \approx 0.45$ (one loop) is identified,
showing that large enough cubic terms increase the roughness. This
fixed point is unstable, revealing a rough strong-coupling
phase. Experimental study of contact angles $\theta$ near $\pi/2$,
where cubic terms in the energy vanish, is suggested.
\end{abstract}
\maketitle

Experiments measuring the roughness of the contact line of a fluid
wetting a disordered solid substrate have consistently found a value
$\zeta \approx 0.5$ for the roughness exponent.  This result is
highly reproducible from superfluid Helium \cite{rolley} to viscous
glycerol-water mixtures \cite{moulinet}, and in situations which can
rather convincingly be argued to be at or at least very near the
depinning transition. Explaining this high value for $\zeta$ poses a
theoretical challenge. It may result in a broader understanding of the
depinning transition in other systems, since similar values
are also measured in cracks \cite{cracks1}.

The simplest elastic model of a contact line \cite{JoannyDeGennes1984}
consists of an effective elastic energy, quadratic in the height field
$h(x)$ (displacement in the solid plane), with non-local dispersion $c
|q|$ (long-range elasticity) due to the surface tension of the fluid
meniscus. Substrate inhomogeneities are modeled by a random-field
disorder coupling to $h(x)$. The resulting model for the depinning
transition of an elastic manifold (generalized to $d$ internal
dimensions, here $d=1$) has been extensively studied, and the
predictions debated for some time.  Functional RG (FRG) methods were
developed initially to one loop \cite{ErtasKardar1994b}
predicting $\zeta=\epsilon/3$ to all orders, here $\epsilon=2-d$,
identical to the statics of random field.
Careful analysis beyond one loop however revealed 
new irreversible terms in the RG which clearly distinguish statics and
depinning, and yield $\zeta=\epsilon/3(1 + 0.397 \epsilon) +
O(\epsilon^3)$ \cite{ChauveLeDoussalWiese}.  Novel high-precision
numerical algorithms found \cite{RossoKrauth} $\zeta=0.388\pm 0.002$ midway
between the one- and two-loop results.  This value is too low to
account for the experiments.

Various mechanisms have been proposed \cite{bouchaud} 
such as lateral waves or plastic-type
dynamics.  It is unclear whether any of them are universal enough to
explain the robustness of the experimental values for 
$\zeta$. More complex dissipation
mechanism may be at play but they should not be important for the
roughness if, as believed \cite{moulinet}, the experiment is at
quasi-static depinning.

Before abandoning the elastic model, one must first check
for neglected effects. It has been known for some time, for 
conventional linear elasticity $c q^2$, that there is
another universality class, anisotropic depinning
\cite{aniso,RossoKrauth2001b,TangKardarDhar1995}.  There, a non-linear
Kardar-Parisi-Zhang (KPZ) term $\lambda (\nabla h)^2$ becomes
relevant, resulting in a singular dependence of the threshold force
on rotation of the contact line in the plane. That such a
dissipative term is generated in the equation of motion at
velocity $v>0$ is straightforward, and it was shown recently within
the FRG that it survives \cite{LeDoussalWiese2002a} even at
quasi-static depinning $v \to 0^+$, only if anisotropy exists in the
substrate disorder or in the motion of the manifold.
Although Ref.~\cite{LeDoussalWiese2002a} shows that a
larger exponent is possible for long-range elasticity, an
explanation based on this term alone is problematic.

An important feature of the contact-line problem
is that at contact angles different from 
$\theta=\pi/2$ the symmetry $h \to -h$ is absent. Thus
the elastic {\it energy} contains cubic non-linear and non-local terms.
The equation of motion thus contains non-linear terms breaking
this symmetry, even in the absence of a driving force \cite{GR}. 
Being different in nature from the conventional
dissipative KPZ term, it is important to understand their effect at
quasi-static depinning. Their effect in
the moving phase has been recently investigated \cite{GR},
and found to lead to a roughening transition
analyzed in connection with the 
development of a Landau-Levich film. 

The aim of this letter is to examine whether non-linear terms in the
energy may result in a different universality class at
depinning. Although naively irrelevant, they do generate at depinning
a {\it non-local} KPZ-type term which grows under coarse-graining.  We find
via a FRG calculation two possible phases separated by a fixed point
at a critical value of the disorder.  Interestingly, the roughness
exponent at this critical point is, within one-loop accuracy, $\zeta
\approx 0.45 >1/3$.  Although we do not control the rough phase at
strong disorder, this shows that $\zeta$ can be increased by
non-linear terms.  This growing non-local term also arises in the
moving phase, with similar features and a significantly
smaller exponent. We discuss the interest of studying the system for
interfaces with contact angles in the vicinity of $\theta=\pi/2$, the
point where cubic terms in the energy vanish.

\begin{figure}[t] \centerline{ \fig{6cm}{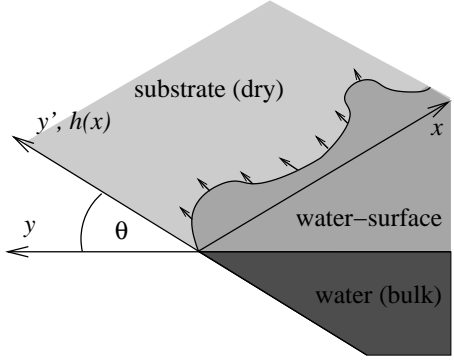} } \caption{The
geometry of the contact-line.}  \label{clgeometry}
\end{figure}
Let us describe the model of Ref.~\cite{JoannyDeGennes1984,GR} for a
fluid wetting a flat solid in coordinates suitable for any (global)
equilibrium contact angle $\theta=\theta_e$, see
Fig.~\ref{clgeometry}.  The liquid-air interface (LA) is denoted by
$\{x,y,z(x,y)\}$ (flat when $z(x,y)=0$), the flat solid (S) surface by
$\{x,y,-y \tan \theta\}$. They meet at the contact line $y = h(x) \cos
\theta$, included in (S), the boundary condition $z(x,y=h(x) \cos
\theta) = - h(x) \sin \theta$.  The total energy is
$E=E_{\mathrm{LA}}+E_{\mathrm{SL}}$ with:
\begin{eqnarray}
E_{\mathrm{\mathrm{LA}}}[z] &=& \gamma_{\mathrm{\mathrm{LA}}} 
\int \rmd x \int_{y > h(x) \cos \theta} \rmd y\,  A[z(x,y)] \\
E_{\mathrm{\mathrm{SL}}} &=& \int \rmd x \int_{y' > h(x)} \rmd y'
\,\gamma(x,y') \ ,
\end{eqnarray}
where $A=\sqrt{1 + (\partial_x z)^2 + (\partial_y z)^2}$ gives the LA
area, and the wetting area energy density $\gamma=\gamma_{\mathrm{SL}}
- \gamma_{\mathrm{AS}}$ is a random function of the in-plane position
($y'=y/\cos \theta$). Minimizing the LA interface energy
$E_{\mathrm{LA}}$ for fixed $h(x)$ yields the equilibrium profile
$z_h(x,y)$, and one can show that the relation between force and
contact angle holds locally, \ $\delta E_{\mathrm{LA}}[z_h]/\delta
h(x) = \gamma_{\mathrm{LA}} \cos \theta(x)$. One defines the
equilibrium contact-angle value $\theta_e$ via the average force,
i.e.\ through $\gamma(x,y') = \gamma_{\mathrm{LA}} \cos \theta_e +
\tilde \gamma(x,y')$ with $\overline{\tilde \gamma(x,y')}=0$.

Expansion of $E_{\mathrm{LA}}[z_h]$ up to third order in $h$ proceeds
by solving $\nabla^2 z_h=0$ in the form $z_h(x,y)=\int_q \alpha_q
e^{iqx - |q|y}$.  Solving for the boundary conditions and inserting in
$E_{\mathrm{LA}}[z_h]$ yields, up to terms of $O(h^4)$:
\begin{equation}
E_{\mathrm{LA}}[h] = \frac{c_1}{2} \int_q |q| h_{q} h_{-q} +
\frac{\lambda}{2} \int_{q,k} [ |q| |k| + q k ] h_k h_q h_{-q-k}
\label{energy}
\end{equation}
with $c_1=\gamma \sin^2 (\theta$), $\lambda = c_1 \cos (\theta) $. 
The form of the cubic term
$e_{q_1,q_2,q_3} h_{q_1} h_{q_2} h_{q_3}$,  $\sum q_i=0$,
with $e_{q_1,q_2,q_3} \sim \sum_{i<j} q_i q_j \Theta(q_i q_j)$
is the only possibility imposing that $e$ is symmetric, homogeneous
of degree 2, and vanishes for $q_1=0$ (invariance under a uniform shift
$h(x) \to h(x) + cst$). As expected the symmetry  $h_k \to - h_k$
is restored for $\theta=\pi/2$ where $\lambda=0$. 
To the same order the general form of the equation of motion (EOM) is:
\vspace*{-.8cm}

\begin{widetext}
\begin{equation}
\eta \partial_t h_{k,t} = - c(k) h_{k,t} - \frac{1}{2} \int_q
\left[ \lambda_2(q,k-q) + \lambda_3(-k,q) + \lambda_3(-k,k-q) \right]
h_{q,t} h_{k-q,t} 
+ \int_x \left[ F(x,h(x)+vt) + f - \eta v\right] e^{-ikx}
\label{eom}
\end{equation}
\vspace*{.0cm}

\end{widetext}
\vspace*{-1.2cm}

\noindent where the pinning force has correlator $\overline{F(x,h) F(x',h')} =
\delta^d(x-x') \Delta(h-h')$, and a thermal noise may be added. In
view of later RG calculations, we slightly generalize the model,
defining (with normalized vectors $\hat q= q/|q|$):
\begin{equation}
\lambda_i(q_1,q_2) = |q_1|^\alpha |q_2|^\alpha f_i(\hat q_1 \cdot \hat q_2) 
\ , \qquad c(q)=c_\alpha |q|^\alpha \label{form1}
\end{equation}
The contact line 
corresponds to $\alpha=1$. For reasons detailed below
we consider the parameterization:
\begin{equation}
f_2(z) = \lambda_3 (1 + g(z)) +  \lambda_0 \ , \qquad 
f_3(z) = \lambda_3 (1 + g(z)) \label{form2}
\end{equation}
where $g(z)=z$ or more generally 
an odd function with $g(-1)=-1$. This EOM is obtained from (\ref{energy}) in the simplest
case, which assumes fast relaxation of the meniscus and dissipation via
molecular jumps \cite{Blake,GR}:
\begin{eqnarray}\label{iiui}
\frac{\eta \partial_t h(x,t)}{\sqrt{1 {+} (\partial_x h)^2}}
&=& \frac{- \delta E[h]}{\delta h(x,t)}\nonumber \\
& =& \gamma [\cos \theta(x,t) 
- \cos \theta_e] + \tilde \gamma(x,h(x,t))
\ ,\qquad
\end{eqnarray}
where $\eta$ is a dissipative coefficient
(note that the above equation neglects viscous hydrodynamic
losses inside the moving liquid wedge  \cite{dG1}).

At zero or vanishingly small velocity, i.e.\ at the depinning
threshold, the non-linearity $(\partial_x h)^2$ on the left-hand side
can be neglected to this order, and using the form (\ref{energy}) one
finds (\ref{eom}) and (\ref{form2}) with $\lambda_3=\lambda$,
$\lambda_0=0$ and $F(x,h)=\tilde \gamma(x,h)$. These are the
microscopic (bare) values, and we show below that the form given in
(\ref{form1}) and (\ref{form2}) is preserved under RG at depinning.
The crucial point is that, while forbidden at the bare level by the
potential form of the EOM $\eta \partial_t h = - \delta E/\delta h$, a
non-zero and positive value for $\lambda_0$ will be generated {\it
beyond the Larkin length} from the non-analyticity of the renormalized
force correlator. It corresponds to the generation of a {\it
non-local} KPZ-type term. The case of more complicated dynamics is
mentioned below. We now focus on the analysis of (\ref{eom}).

\begin{figure}[b]
\centerline{ \fig{4.5cm}{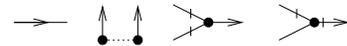} } \caption{The graphical rules: Propagator, 
disorder vertex, non-linearity $\sim \lambda_{2}$
and non-linearity $\sim \lambda_{3}$.}
\label{rules}
\end{figure}%
\begin{figure}[t]
\centerline{$\diagram{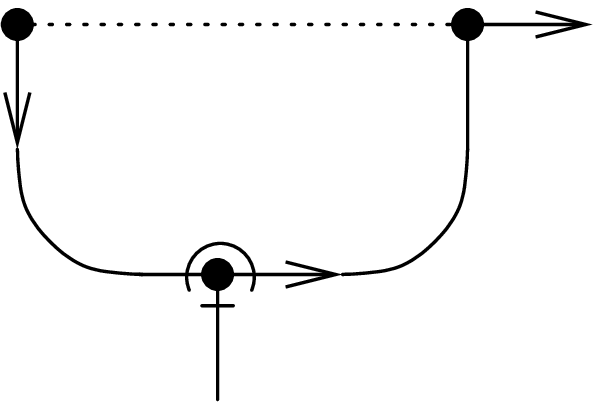} + \diagram{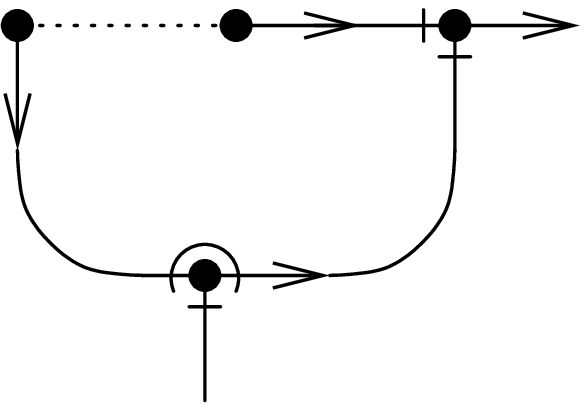}$}
\centerline{$\diagram{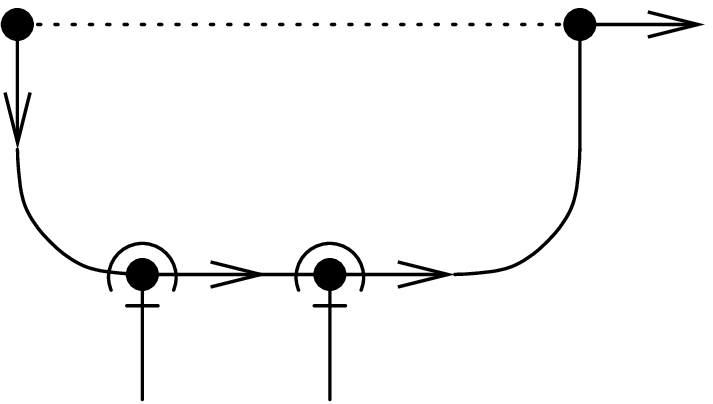}+ \diagram{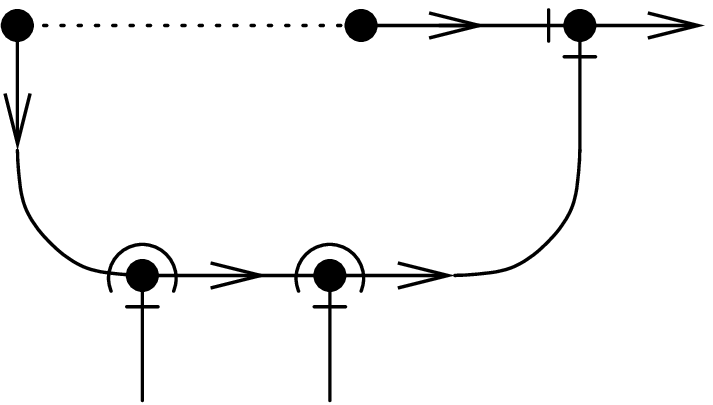}+\diagram{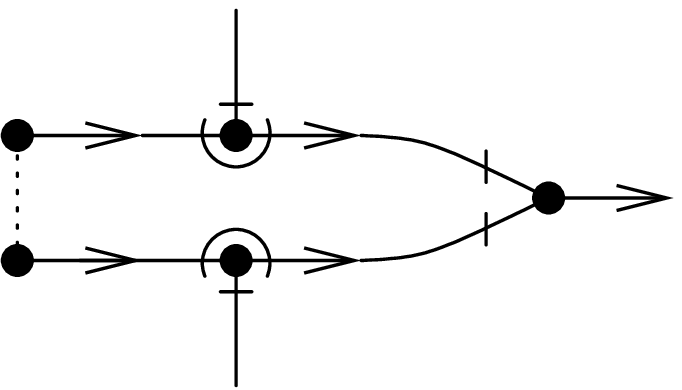} $}
\centerline{$\diagram{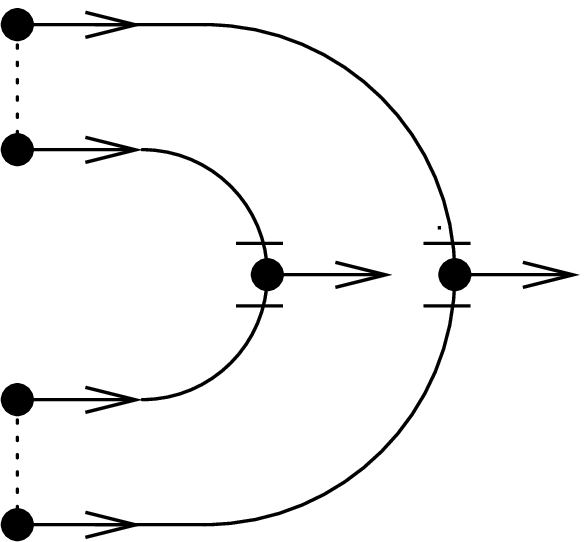} + \diagram{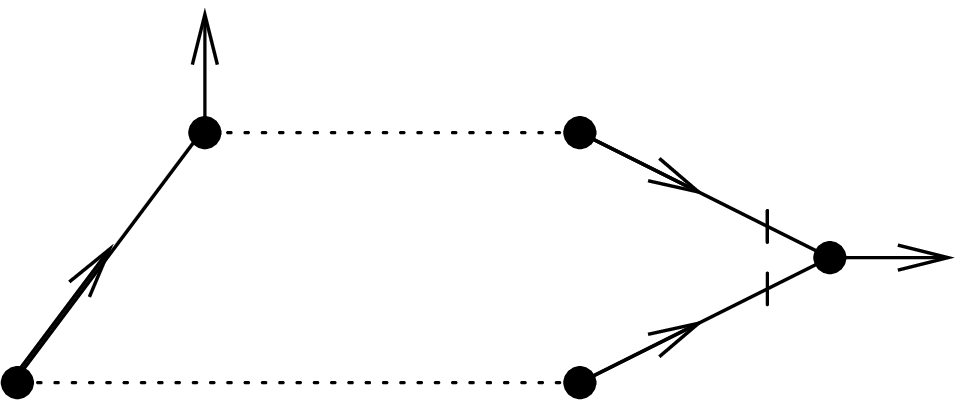}+\diagram{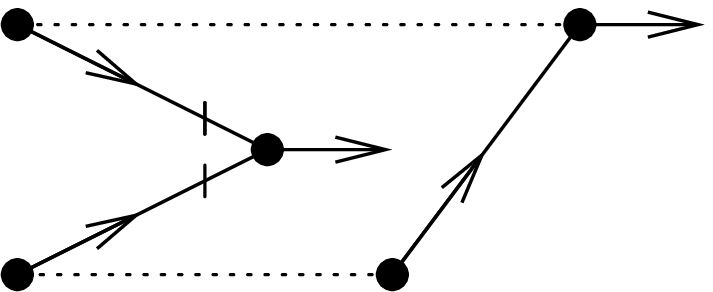} $}
\centerline{$ \diagram{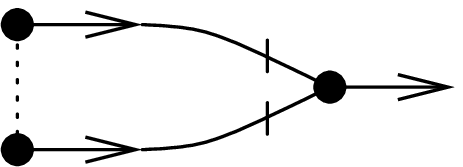} $}
 \caption{The diagrams correcting $c_{\alpha}$ (first line), 
$\lambda_{i}$ (second line), $\Delta(h)$ (third line) and 
$\eta$ (last line). Only diagrams proportional to $\lambda_{i}$ are shown.}
\label{diagrams}
\end{figure}
\begin{figure}[b]
\centerline{$\diagram{KPZ3cr} = \diagram{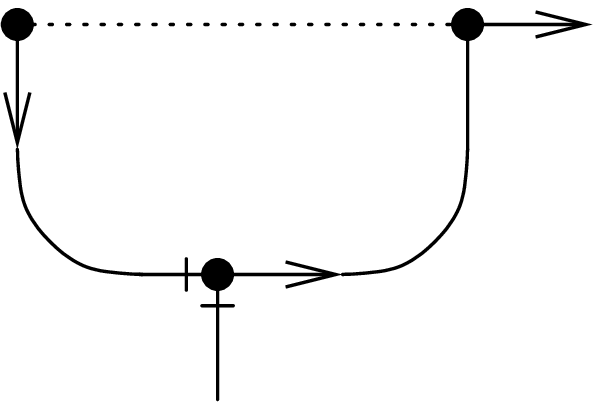} +\diagram{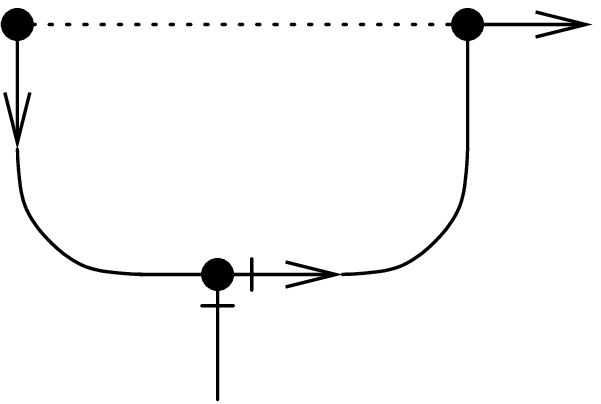} $}
 \caption{The first diagram from Fig.~\ref{diagrams} correcting $c_{\alpha}$, see (\ref{eight}).}
\label{calpha}
\end{figure}

FRG starts by calculating the 1-loop corrections to (\ref{eom}) and
its associated dynamical action, perturbatively in $\Delta(h)$ and the
$\lambda_i$. Graphical rules are given in Fig.~\ref{rules}. The graphs
representing the corrections to $c_\alpha$, $\lambda_i$, $\Delta(h)$
and $\eta$ are shown in Fig.~\ref{diagrams}. To illustrate the main
ideas, we explain how the first correction to $c_\alpha$ is
computed. It is the sum of two contributions (see Fig.~\ref{calpha}):
\begin{equation}\label{eight}
\delta c(p) = \frac{- \Delta' (0^{+})}{c_\alpha^2} \int_k 
\frac{|k|^\alpha |p|^\alpha}{|k+p|^\alpha |k|^\alpha} f_2(\hat k \cdot \hat p) 
+ \frac{|p|^\alpha}{|k|^\alpha} f_3(\hat p \cdot \hat{q})
\ ,
\end{equation}
where $q=-(p+k)$.  These are depicted on Fig.\ \ref{calpha} in the
order of their appearance in (\ref{eight}).  Denominators originate
from time integrals of the bare response $R_{k,t}=(\eta)^{-1}
e^{-c_\alpha |k|^\alpha t/\eta} \Theta(t)$, while numerators come from
the cubic vertices. The factor $\Delta' (0^{+})$ is $\Delta'
(v(t-t'))$, taken in the limit of $v\to 0^+$ and is non-vanishing only
when the cusp is formed, i.e.\ beyond the Larkin length. Expanding in
vanishing external momentum $p$ yields:
\begin{equation}
\delta c_\alpha = - \frac{\Delta' (0^{+})}{c_\alpha^2}  \int_k 
\frac{1}{ |k|^\alpha} \left< f_2(z) + f_3(-z) \right> 
\ ,
\end{equation}
where $\left< f(z)\right>$ denotes the angular average $\sim
\int_{-1}^{1} \rmd z\, (1-z^2)^{\frac{d-3}{2}} f(z) $ (with $\left< 1
\right>=1$). Appearance of the combination $\left<f_2(z) +
f_3(-z)\right> = 2 \lambda_3 + \lambda_0$ for arbitrary odd $g(z)$ in
all graphs is a general feature and yields the (re)definition of the
vertex explained in Fig.\ \ref{calpha}. It gives the ``anomalous''
correction (i.e.\ resulting from the cusp) to $c_\alpha$ to which the
second one of Fig.~\ref{diagrams} must be added.
One shows that there is no correction to the non-linear vertex
function $f_3(z)$ because that would involve necessarily another $f_3$
vertex with an external $h$ leg at zero external momentum, yielding
$f_3(-1)=0$ (which remains true since there is no mechanism to correct
it).  Thus one finds $\delta \lambda_3=0$. In the graphs correcting
$f_2$ (second line of Fig.~\ref{diagrams}) appearance of the
above-mentioned combination implies that only the uniform part of
$f_2$, i.e.\ $\lambda_0$ is corrected:
\begin{equation} 
\delta \lambda_0 = (2 \lambda_3 + \lambda_0)^2 \left[ - 2 \frac{\Delta'
 (0^{+})}{c_\alpha^3} \int_k k^{-\alpha}
+ 3 \frac{\Delta (0)}{c_\alpha^4} \lambda_0 \int_k \right]
\ .
\end{equation}
While in standard (thermal) KPZ the two corresponding diagrams have
opposite sign because of the derivatives and $\lambda$ is uncorrected
(a consequence of Galilean invariance), here the presence of {\it
absolute values} of the momenta results in the same sign.

Corrections to disorder (line 3 of Fig.~\ref{diagrams})
are the same as in \cite{LeDoussalWiese2002a}. Power counting then 
leads to the definitions $\tilde \Delta(h) = \Lambda_l^{d - 2 \alpha+ 2 \zeta} c_\alpha^{-2}
 \Delta(h \Lambda_l^{-\zeta})$,
$\tilde \lambda_i = \Lambda_l^{\alpha
 - \zeta} c_\alpha^{-1} \lambda_i$, 
where $\Lambda_l=\Lambda e^{-l}$ is the
running UV cutoff (e.g.\ in a Wilson approach).
One finally obtains the following set of FRG equations
(with $\epsilon=2 \alpha -d$):
\begin{eqnarray} \label{lf96}\nonumber 
  \partial_l \ln c_\alpha &=& \tilde \Delta' (0^{+}) (2 \tilde
  \lambda_3 + \tilde \lambda_0) - \tilde \Delta(0)
(2 \tilde \lambda_3 + \tilde \lambda_0 ) \tilde \lambda_0 
 \\
\nonumber 
  \partial_l \ln \tilde \lambda_3 &=& \zeta - \alpha -  \partial_l \ln c_\alpha \\
\partial_l \tilde \lambda_0 &=& (\zeta - \alpha - \partial_l \ln
c_\alpha) \tilde \lambda_0
- 2 \tilde \Delta' (0^{+}) (2 \tilde \lambda_3 + \tilde \lambda_0)^2 
\nonumber 
\\ &&+ 3 \tilde \Delta (0) \tilde \lambda_0 (2 \tilde \lambda_3 + \tilde
  \lambda_0 )^2  
\\ \nonumber 
\partial_l \tilde \Delta(h) &=& (\epsilon  - 2 \zeta - 2 \partial_l
\ln c_\alpha) \tilde \Delta(h)
+ \zeta h \tilde \Delta'(h) \\
\nonumber 
&&+  \frac{1}{2} \tilde \Delta(h)^2 \tilde
  \lambda_0^{2} - [ \tilde \Delta'(h)^2 + \tilde \Delta''(h)( \tilde \Delta(h) -
\tilde \Delta(0) )] \nonumber \\
 \partial_l \ln \eta &=& \frac{1}{2} \tilde \Delta' (0^{+}) 
 \tilde \lambda_0 - \tilde \Delta'' (0^{+}) 
\ .\nonumber 
\end{eqnarray}
They admit the standard attractive depinning fixed
point corresponding to linear elasticity (isotropic IS
depinning class): $\tilde \lambda_3=\tilde \lambda_0=0$ and $\tilde
\Delta_{\mathrm{IS}}^*(h)$ with $\zeta=\epsilon/3$ to this order
(corrected at two loops). To order
$\epsilon$, this FP is stable to adding a small non-zero $\lambda_3,
\lambda_0$ which have linear eigenvalue $\zeta - \alpha$. This FP
controls a phase with small $\lambda_{0}$ and
$\lambda_{3}$. $\lambda_0$ is generated from $\lambda_3$ beyond the
Larkin length and the ratio $\lambda_0/\lambda_3$ goes to a constant
in this phase.

We found a second fixed point which controls at given disorder the
transition between the small $\tilde \lambda_0$ phase and a large
$\tilde \lambda_0$ regime (strong coupling).  One easily sees that the
ratio $\tilde \lambda_3/\tilde \lambda_0$ flows to zero at this
transition. To look for the FP one can thus set $\tilde \lambda_3=0$
and redefine $\hat \Delta(h) = \lambda_0^2 \tilde
\Delta(h/\lambda_0)$, yielding:
\begin{eqnarray} \label{m91a}
 \partial_l \hat \Delta(h) &=& [\epsilon - 2 \hat \zeta - 2 
\hat \Delta' (0^{+}) + 2 \hat \Delta(0) ]  \hat \Delta(h)
+ \hat \zeta h \hat \Delta'(h) \nonumber \\
&& + \frac{1}{2} \hat \Delta(h)^2 - [ \hat \Delta'(h)^2 + \hat
\Delta''(h)( \hat \Delta(h) - \hat \Delta(0) )] 
\nonumber\\
\end{eqnarray}
with $\partial_l \ln \tilde \lambda_0 = \zeta - \hat \zeta$
and $\hat \zeta= \alpha + 3 \hat \Delta'(0^+) - 4 \hat \Delta(0)$.
The FP function obtained numerically $\hat \Delta^*(h)=\epsilon f_a(h)$ 
is positive and short-ranged, and $f_a$
depends only on $a=\alpha/\epsilon$; the associated 
roughness exponent is
\begin{eqnarray} \label{m91b}
 \zeta &=& 0.450512 \epsilon 
\end{eqnarray}
for $a=1$, which should be compared to the value $\zeta_{\mathrm{IS}}
= \epsilon/3$ to the same 1-loop accuracy, demonstrating the increase
in the roughness exponent due to non-linearities. If we assume the
same relative increase of $\zeta$ due to the higher-loop corrections
\cite{ChauveLeDoussalWiese} from $1/3$ to the observed
$\zeta_{\mathrm{IS}} \approx 0.388 \pm 0.002$ \cite{RossoKrauth}, we would
obtain here $\zeta \approx 0.53$, tantalizingly close to the
experiments. The dynamical exponent is $z =\alpha + \partial_l
\ln(\eta/c)$,
i.e.\ $z=1 - 0.205213 \epsilon$ yielding $z=0.79487$ for
the physical case. As for anisotropic depinning, a third exponent is
necessary here, $\psi = - \partial_{\ell } \ln c_\alpha = 0.170449 \epsilon$,
and scaling yields the correlation length exponent
$\nu=1/(\alpha-\zeta+\psi)$ and the velocity-force exponent  
$\beta=\nu(z-\zeta) \approx 0.478$ such that $v \sim (f-f_c)^\beta$.
This fixed point is unstable in one direction (leading eigenvalues
$\mu_1=0.938$ and $\mu_2=-1.23$) consistent with the existence of two
phases. The strong-coupling phase cannot be accessed by the present
method, but the increase of $\zeta$ is likely to
persist there. Since this FP is attractive in all other directions 
the experiments may be susceptible to a very
long crossover dominated by this FP. This can be tested by careful
numerical integration of (\ref{lf96}), not attempted
here.

The FRG equations (\ref{lf96}) have been derived within a double
expansion in $\epsilon$ and $\alpha$. The question arises of whether
operators with more non-local derivatives $|\nabla |^{\alpha}h \sim
|k|^\alpha h$ are indeed irrelevant, as suggested by power-counting. A
detailed analysis shows that in the space of perturbations where one
adds to $\Delta (h_{xt}-h_{xt'})$ the two functions $\Delta_{2s}
(h_{xt}-h_{xt'})\left(|\nabla|^{\alpha} h_{xt}+|\nabla|^{\alpha}
h_{xt'} \right)+ \Delta_{2u} (h_{xt}-h_{xt'})\left(|\nabla|^{\alpha}
h_{xt}-|\nabla|^{\alpha} h_{xt'} \right) $, the largest eigenvalue is
$-0.12$ (for $\epsilon =\alpha =1$), indicating no additional
instability of the FP. These terms arise at the bare level
for a more complicated dynamics, but should not change the result in the
quasi-static limit studied here \cite{us}.

We now sketch the analysis of the moving case, very near depinning,
$v>0$ small, for details see \cite{us}. Since the generation
of $\lambda_0$ is a new feature, we reexamine \cite{GR}. 
At large scales in the moving phase the quenched pinning
force acts as a (thermal) white noise of strength $2 D$ (notation as
in \cite{GR} ) and thus $\eta$ is uncorrected. Using the same
parameterization as above we find \cite{footnote}:

\begin{eqnarray} 
\partial_l \ln \tilde c_\alpha &=& z - \alpha - \frac{1}{4} g (1+2 r)
\nonumber 
\\
 \label{P153} \partial_l \ln \tilde \lambda_0 &=& z + \zeta - 2 \alpha +
\frac{1}{4} g (1+2 r)^2  \\ \partial_l g &=& - (d+
\alpha) g   + g^2 \Big[\frac{1}{8} + \frac{3}{4} (1+2 r) + \frac{1}{2}
(1+2 r)^2 \Big]\nonumber 
\end{eqnarray}

\noindent 
with $\partial_l \ln r = - \frac{1}{4} g (1+2 r)^2$ and we have defined 
$\tilde c_\alpha= c_\alpha \Lambda_l^{\alpha - z}$,
$\tilde \lambda_i = \lambda_i \Lambda_l^{2 \alpha - (z+ \zeta)}$
($i=0,3$), 
$\tilde D  =  D \Lambda_l^{d+ 2 \zeta -z}$,
$g = \tilde c_\alpha^{-3} \tilde D  \tilde \lambda_0^2$, and
$r = \tilde \lambda_3/ \tilde \lambda_0$. These
equations exhibit a weak-coupling phase controlled 
by the $g=0$ attractive fixed point (which, in
the physical case corresponds to logarithmic roughness
$\zeta=0$) and a strong-coupling phase. They are separated
by a FP at $g=g^*=\frac{8}{11} (d+ \alpha)$, with {\it universal}
values for the exponents
$z=\alpha+\frac{2}{11} (d+ \alpha)$ and 
$\zeta= \frac{\alpha - d}{2} + \frac{3}{11} \frac{d+ \alpha}{2}$,
i.e.\ for $\alpha =d=1$:
\begin{eqnarray}
 \zeta \approx 0.273 \quad , \quad z \approx 1.364\ . 
\end{eqnarray}
This rather low value for $\zeta$ suggests that a scenario based on a
slowly moving contact line is not adequate to explain the
experimentally observed roughness.

In conclusion, we have examined using RG the effect of non-linear
elasticity for the contact line at and near depinning.  We found that
even for isotropic disorder, a non-local KPZ term is generated and
may, for large enough bare non-linear elasticity and disorder,
destabilize the standard linear-elasticity depinning fixed point,
yielding values for the roughness exponent compatible with
experiments. This scenario could be tested by high-precision numerics.
It may also be explored experimentally by carefully choosing the
fluid and the solid substrate \cite{DGBWQ}: since all odd non-linear
terms in the elastic energy vanish at $\theta=\pi/2$ one can surmise 
that the total effect of non-linear terms, and hence 
the apparent contact-line roughness, is
minimal there \cite{footnotetheta}.


\begin{thebibliography}{999}
\bibitem{footnote} Additional non-potential terms appearing at $v>0$
are found to be irrelevant (as compared to $\lambda_0$) if $v$ is
small enough, hence (\ref{P153}) is restricted to small $v$.
\bibitem{footnotetheta}
Perturbative calculations \cite{us} show that even for
$\theta=\pi/2$ a smaller but positive $\lambda_0$ term
is generated from fourth order non-linear elasticity
beyond the Larkin Length.
\end{thebibliography}

We thank C.\ Guthmann, W.\ Krauth,  S.\ Moulinet, E.\ Rolley, A.\
Rosso and T.\ Vilmin for interesting discussions.

\end{document}